# Generalizing Monin-Obukhov similarity theory (1954) for complex atmospheric turbulence

Ivana Stiperski[a] and Marc Calaf[b]

[a] Department of Atmospheric and Cryospheric Sciences, University of Innsbruck, 6020 Innsbruck, Austria; [b] Department of Mechanical Engineering, University of Utah, Salt Lake City, UT 84112, USA

**Monin-Obukhov similarity theory (MOST) stands at the center of understanding of atmospheric surface-layer turbulence. Based on the hypothesis that a single length scale determined by surface fluxes governs the turbulent exchange of momentum, mass, and energy between the Earth's surface and the atmosphere, MOST has been widely used in parametrizations of turbulent processes in virtually all numerical models of atmospheric flows. However, its inherent limitations to flat and horizontally homogeneous terrain, the non-scaling of horizontal velocity variances, and stable turbulence have plagued the theory since its inception. Thus, the limitations of MOST, and its use beyond its intended range of validity contribute to large uncertainty in weather, climate, and air-pollution models, particularly in polar regions and over complex terrain. Here we present a first generalized extension of MOST encompassing a wide range of realistic surface and flow conditions. The novel theory incorporates information on the directionality of turbulence exchange (anisotropy of the Reynolds stress tensor) as a key missing variable. The constants in the standard empirical MOST relations are shown to be functions of anisotropy, thus allowing a seamless transition between traditional MOST and its novel generalization. The new scaling relations, based on measurements from 13 well-known datasets ranging from flat to highly complex mountainous terrain, show substantial improvements to scaling under all stratifications. The results also highlight the role of anisotropy in explaining general characteristics of complex terrain and stable turbulence, adding to the mounting evidence that anisotropy fully encodes the information on the complexity of the boundary conditions.**

Atmospheric Boundary-Layer | Similarity Scaling | Turbulence Anisotropy

Accuracy in numerical weather prediction (NWP) and climate models is limited by three major challenges: the accurate representation of physical processes, sensitivity to model initialization, and the chaotic nature of the atmosphere requiring the use of ensemble forecasting (1). In this work, we focus on the first challenge. Specifically, we generalize the seventy-year old Monin-Obukhov Similarity Theory (MOST, 2) used in practically all NWP, ocean and climate models to parametrize the surface-atmosphere turbulent exchange (3), and extend its applicability from flat and horizontally homogeneous terrain (canonical conditions), for which it was developed, to all surface-atmosphere conditions.

Turbulence is the last unsolved problem in classical mechanics (4). Playing a pivotal role in the climate system, turbulence is the primary mechanism coupling the hydrosphere, biosphere, cryosphere and the atmosphere (5, 6). At the same time, it is the largest source of uncertainty in the Earth's energy budget (7). On shorter timescales, turbulence controls the dynamics of atmospheric phenomena as diverse as tropical cyclones (8), downslope windstorms (9), terrestrial carbon uptake (10), and dispersal of pollutants, nuclear fall out (11) or pathogens (12). Yet, the lack of understanding and accurate representation of turbulence in all numerical models remains a major source of uncertainty, challenging accurate long-term weather prediction and climate projections (1, 13, 14).

Dynamically, the atmosphere can be grossly structured into two distinct regions. An upper region, referred to as the free atmosphere, is controlled by the large-scale differential heating, the gravitational force, and the earth's rotation (15), and is not influenced by the roughness of the earth surface. In the free atmosphere, the flow dynamics are to a good approximation laminar and quasi-two-dimensional, and described well by a closed system of differential equations. Below the free atmosphere, a near surface layer directly influenced by the earth's surface is called the Atmospheric Boundary Layer (ABL, (16)). In this region the flow dynamics are controlled by surface forcing (surface roughness, heating and cooling), the flow is three-dimensional and turbulent, and the system of differential equations describing it remains unclosed. To make the system integrable (13, 16), parametrizations in the form of statistical approximations of turbulent exchange are required, which can be used in numerical weather prediction, ocean (17), and climate models (13). The random nature of turbulence and wide range of scales over which it acts, however, require these parametrizations to rely on imperfect stochastic closure techniques, thus adding to the challenge of complexity of non-linear systems (18), and significantly limiting the

**Significance Statement**

Monin-Obukhov Similarity Theory (MOST) is the cornerstone of atmospheric turbulence and forms the basis for parametrizations of turbulent exchange in virtually all numerical models of atmospheric and oceanic flows. Its limitations to flat and horizontally homogeneous terrain and non-scaling of horizontal momentum variances have, however, plagued the theory since its inception. Here we present a first generalized extension of similarity theory universally valid in conditions in which MOST fails. This new breakthrough theory allows a correct representation of turbulence effects in weather, climate, and air-pollution models over the majority of Earth's surface, thus reducing uncertainties in the surface-atmosphere coupling, especially over particularly vulnerable regions such as polar regions and mountains experiencing unprecedented warming.

predictability of atmospheric flows (19). Understanding and correctly representing turbulence therefore poses a paramount challenge in atmospheric sciences.

Traditionally, the non-linear turbulent exchange of momentum, heat, and mass between the earth surface and the overlying atmosphere has been parametrized through MOST (2). Given its critical role in explaining the near surface turbulent exchanges, MOST has become one of the most celebrated theories in ABL of the last seventy years (20). Based on dimensional analysis (16, 21), MOST states that in the absence of subsidence (mean vertical flow), in steady state conditions (weak time dependence of turbulence statistics), in the near-surface layer characterized by semi-constancy of turbulence fluxes with height, and over horizontally homogeneous and flat surfaces, a single length scale, the Obukhov length ($L$) fully characterizes all surface-atmosphere exchange processes. This length scale $L = -u_*^3 \overline{\theta}/\kappa\, g\overline{w'\theta_v'}$ (22) is obtained from a reduced set of variables: the surface friction velocity ($u_* = (\overline{u'w'}^2 + \overline{v'w'}^2)^{1/4}$) as a fundamental turbulent velocity scale characterizing shear production of turbulence; the surface kinematic sensible heat flux ($\overline{w'\theta_v'}$) quantifying the buoyancy production of turbulence; the buoyancy parameter ($g/\theta$); and the von Kármán constant $\kappa$, approximately equal to 0.4. Here $\overline{u'w'}$ and $\overline{v'w'}$ are the streamwise and spanwise momentum flux. As a result, any mean quantity $x$ in the surface layer (lowest 10% of the ABL where turbulent fluxes are assumed to be quasi-constant with height) when properly non-dimensionalized with the respective turbulent scaling variable $x_*$ is as a universal function of a single scaling parameter $\zeta = z/L$ only (16):

$$\frac{x}{x_*} = \Phi_x(\zeta). \qquad [1]$$

Here $z$ is the height above the surface, whereas examples of turbulence scaling variables $x_*$ include friction velocity $u_*$, and temperature scale $\theta_* = \overline{w'\theta_v'}/u_*$. MOST itself does not provide the functional form of the scaling relations $\Phi_x$ for different variables, instead, these have been experimentally determined (23) and refined (24–27), and as such are widely used in parametrizations.

Still, MOST suffers from significant failures that limit its applicability (20). Some of the most salient limitations are the lack of scaling of horizontal (streamwise and spanwise) momentum variances under unstable thermal stratification (e.g., 28, 29), the non scaling of surface-normal velocity and temperature variances in stable stratification (30), as well as the general breakdown of scaling for intermittent turbulence (31, 32), where the presence of so-called submeso motions requires alternative approaches (33, 34). The deceptive simplicity and general success of MOST have, however, led to its widespread use even under flow and surface conditions where its assumptions are clearly violated. Thus, MOST is routinely applied in NWP over complex terrain (i.e. heterogeneous landscapes and mountainous topography), despite consistent evidence that scaling relations are no longer universal in non-canonical terrain and MOST generally fails there (e.g., 35–43).

Alternative scaling approaches have been suggested to address some of these shortcomings of MOST. These include the mixing layer approach for scaling of horizontal momentum variances in convective conditions (44, 47, 48), requiring the inclusion of a non-local length scale (depth of the convective boundary layer); the local scaling approach in stable stratification (49–51), in which turbulence is scaled with values at the same measurement level; and z-less scaling in very stable stratification, in which turbulence is assumed to be detached from the surface and thus no longer depends on height. All these and other approaches (e.g., 52–54), however, address only individual shortcomings of MOST in a limited set of conditions (specific surface type or flow conditions).

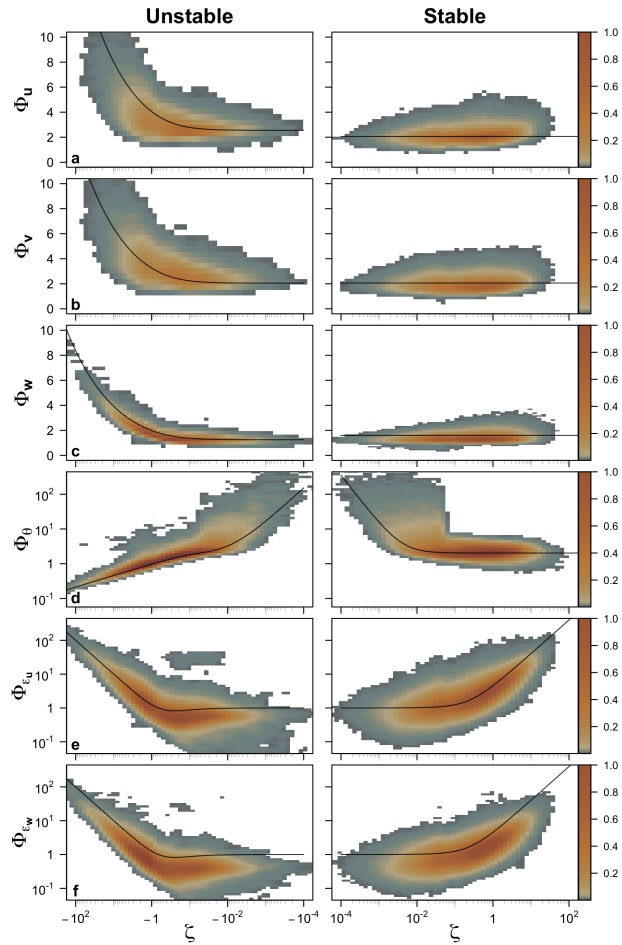

**Fig. 1.** Two-dimensional density plots of flux-variance scaling relations for standard deviations of a) streamwise ($\Phi_u$), b) spanwise ($\Phi_v$) and c) surface-normal ($\Phi_w$) momentum, d) temperature ($\Phi_\theta$), and e) streamwise ($\Phi_{\varepsilon_u}$) and f) surface-normal ($\Phi_{\varepsilon_w}$) TKE dissipation rate, as a function of non-dimensional scaling parameter $\zeta$, for unstable (left) and stable (right) stratification. Shown is the ensemble of all datasets used in the study. Black line corresponds to traditional scaling curves corresponding to horizontal homogeneous and flat terrain for unstable and stable stratification respectively (16, 42, 44–46): a) $\Phi_u = 2.55(1-3\zeta)^{1/3}$, $2.06(\zeta)^0$, b) $\Phi_v = 2.05(1-3\zeta)^{1/3}$, $2.06(\zeta)^0$, c) $\Phi_w = 1.35(1-3\zeta)^{1/3}$, $1.6(\zeta)^0$, d) $\Phi_\theta = 0.99(0.067 - \zeta)^{-1/3}$ for ($\zeta < -0.05$) and $0.15(-\zeta)^{-1} + 1.76$ for ($\zeta > -0.05$), $0.00087(\zeta > -0.05)^{-1.4} + 2.03$, e) and f) $\Phi_\varepsilon = (1-3\zeta)^{-1} - \zeta$, $(1 + 4\zeta + 16(\zeta)^2)^{1/2}$.

Recently Stiperski et al. (55, 56) proposed a more universal approach that addresses most of the shortcomings and limitations of MOST over a wide range of stratification and surface conditions. Their results showed that the departures from MOST scaling are strongly correlated to the anisotropy of the Reynolds stress tensor (57). If clustered according

2 | Stiperski *et al.*

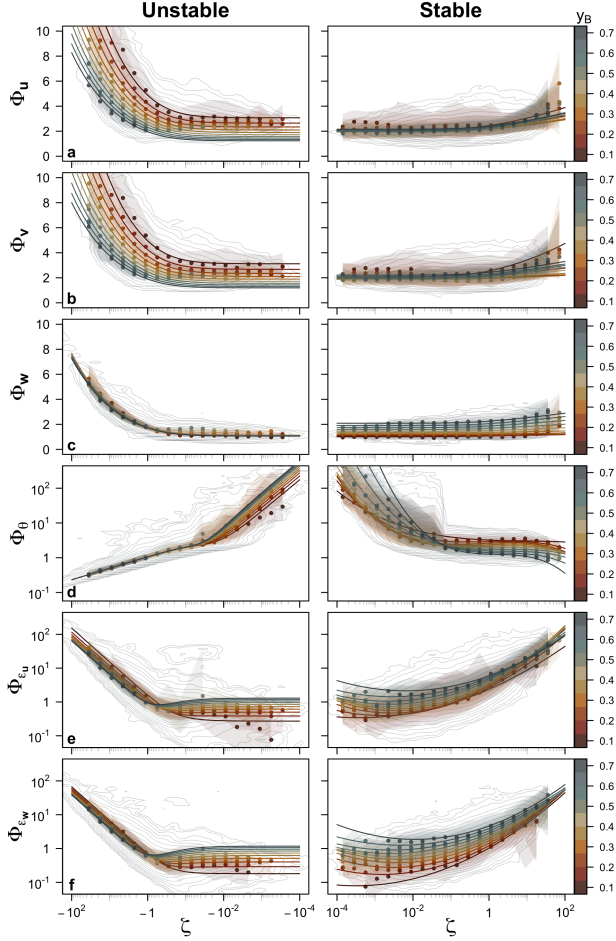

**Fig. 2.** Novel scaling relations for the standard deviations (full lines) of a) streamwise ($\Phi_u$), b) spanwise ($\Phi_v$), and c) surface-normal ($\Phi_w$) momentum, d) temperature ($\Phi_\theta$), and e) streamwise ($\Phi_{\varepsilon_u}$) and f) surface-normal ($\Phi_{\varepsilon_w}$) TKE dissipation rate, as a function of non-dimensional scaling parameter $\zeta$, for unstable (left) and stable (right) stratification, for all datasets clustered according to the degree of anisotropy $y_B$ (color). Points are bin medians and shading 10% to 90% percentiles for each cluster. Gray contours represents the density of the data points from Fig. 1. The definition of the scaling curves is given in Table 1.

to turbulence anisotropy, datasets representative of widely diverse flow and topographic forcing conditions that traditionally do not follow MOST scaling, were shown to collapse onto the same scaling curve. This collapse occurred regardless of the complexity of the terrain or background forcing, and without the need for the addition of a non-local scale as in the mixed-layer scaling (58), showing that turbulence anisotropy is the missing ingredient that encodes the local flow complexity. The new approach was therefore shown to hold beyond canonical conditions which satisfy MOST assumptions, i.e. in complex terrain (56). That the anisotropy is of fundamental importance when characterizing turbulent exchange was further emphasized by Stiperski et al. (59). In this work it was shown that the scalewise path turbulence follows through the energy cascade towards isotropy at the smallest turbulence scales (60) is universal (independent of terrain or flow complexity, or even stratification) and solely dictated by the bulk anisotropy. Both of these results can be explained only if locally computed turbulence anisotropy encodes all relevant information on the boundary conditions affecting the turbulent flows, including the non-local influences. It is hence clear that anisotropy needs to be accounted for in a generalized scaling approach, as already hinted by Kader and Yaglom (29). In this work, we therefore present the first generalized scaling framework for complex atmospheric turbulence, based on turbulence anisotropy. The approach does not supersede, but instead expands upon traditional MOST, which remains the cornerstone of physical process representation in the ABL. When developing the novel generalized scaling, we use a broad selection of 13 turbulence measurement datasets previously published in the literature that represent a wide range of stratifications, and vastly different surface and flow conditions: from flat and homogeneous terrain (CASES-99 (61), AHATS (62), CABAUW (63)) to moderately (T-REX (64), METCRAX II (65)) and highly complex alpine terrain (i-Box (41)). The novel scaling relations not only offer a solution to a seventy-year old problem in ABL meteorology but also allow a deeper understanding of turbulence and its role in earth-atmosphere exchange over realistic terrain, to constrain the uncertainties in the future climate scenarios.

### Flux-variance scaling relations

MOST provides scaling for a range of variables: from gradients, variances to spectra (16). Here we focus on the flux-variance relations that are key to correctly modeling turbulent dispersion (66), as well as understanding the dynamics of stable boundary layer (67). In fact, limited studies (e.g., 42, 43, 46) have indicated that terrain and flow complexity modify turbulent variances differently compared to turbulent fluxes, thus causing a mismatch in scaling relations between the different datasets. The associated large scatter, caused by both the failure of MOST over canonical conditions and due to surface complexity, is clearly visible when plotting traditional MOST flux-variance scaling for the ensemble of datasets with varying degree of complexity (Fig. 1).

**Momentum variance scaling.** As commonly observed (e.g., 36, 68), the scatter of the scaled horizontal momentum variances $\Phi_{u,v}$ differs considerably from scaled surface-normal momentum variance $\Phi_w$. Since surface-normal momentum variance adapts rapidly to the local surface characteristics (Fig. 1c), $\Phi_w$ has small scatter over the different surfaces. The scaled horizontal variances $\Phi_{u,v}$, however, are characterized by substantial scatter, especially in moderately to strongly convective cases (Fig. 1a-b). Stiperski et al. (56) showed that this large scatter can be explained by differences in anisotropy. In fact, Fig. 2a shows that the more isotropic the near surface turbulence tends to be (larger $y_B$ - i.e. anisotropy invariant used to quantify the degree of turbulence anisotropy (69)), the closer the horizontal variances scale to the scaled surface-normal momentum variance. To account for this, we therefore introduce the degree of anisotropy $y_B$ as an additional independent, non-dimensional group in the original similarity theory

$$\frac{x}{x_*} = \Phi_x\left(\zeta, y_B\right). \quad [2]$$

Here $x$ are the standard deviations of the streamwise ($\sigma_u$), spanwise ($\sigma_v$), and surface normal ($\sigma_w$) momentum, and $x_*$ is the friction velocity $u_*$, and the novel empirical scaling relations $\Phi_x$ for each variable and stratification (Table 1), derived in Sect. 1, are illustrated in Fig. 2. The striking feature of the novel curves is their clear stratification according to anisotropy. The clearest dependence on anisotropy, and therefore the largest improvement brought by its inclusion, is observed for variables known not to follow MOST: horizontal momentum variances in convective conditions and surface-normal momentum variance in stable stratification. Importantly, the generalized scaling retains the same functional form as traditional MOST, including its accepted theoretical limits: free convective limit (1/3 power law) in very unstable conditions (29) and z-less limit in stable stratification up to $\zeta = 2$. Nonetheless, the constants in traditional MOST are shown to be functions of anisotropy in the generalized scaling, allowing a seamless transition between traditional MOST and the novel formulation.

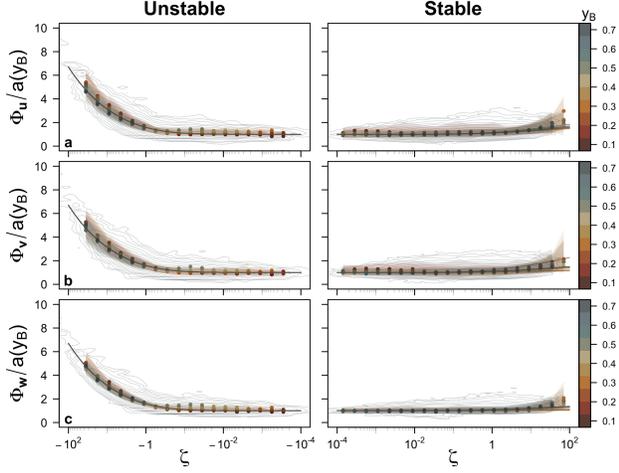

**Fig. 3.** Same as Fig. 2a-c, but normalized by the coefficient $a(y_B)$ (cf. Table 1) as a function of scaling parameter $\zeta$ clustered according to the degree of anisotropy (color). Gray contours represents the density of the data points. Full lines are the corresponding scaling curves.

Given that the dependence on anisotropy is only present in the terms corresponding to the near-neutral limit (i.e., value of scaling curves at the limit of $\zeta = 0$ corresponding to $a(y_B)$ in Table 1), it is possible to define new scaled momentum variances $\Phi_{u,v,w}/a(y_B)$ that are functions of $\zeta$ only (Fig. 3). This new normalization leads to the collapse of all data onto a single curve with a minimal scatter (cf. Fig. 1), a sign of a successful scaling (16). In fact, the scatter of the horizontal momentum variances under the generalized scaling is comparable to the MOST scaling of the surface-normal momentum variance accepted to follow MOST even in complex conditions (36). Additionally, the new normalization outperforms the mixed layer scaling approach commonly assumed to explain the horizontal momentum variances (44) (not shown). This is true for all datasets irrespective of their complexity.

**Temperature and dissipation scaling.** Another salient characteristic of the new generalized approach is the improvement to the scaling of temperature variance ($\Phi_\theta$) and TKE dissipation rate ($\Phi_\varepsilon$). Here $x$ in Eq. 2 are the standard deviation of temperature ($\sigma_\theta$), and the dissipation rate ($\varepsilon$), while the scaling variables $x_*$ are the temperature scale $\theta_*$ and $u_*^3/kz$ respectively. Whereas anisotropy is a characteristic of the Reynolds stress tensor and can be expected to influence momentum scaling, we show that anisotropy plays an important role in scaling of scalars as well. Compared to the scaling of momentum variances (Fig. 1a-c), however, the classical scaling of temperature and dissipation rate (Fig. 1d-f) is highly non-linear (45, 70, 71), and the scaled variables extend over several orders of magnitude. The spread between the datasets and inter-dataset scatter is also significant, highlighting the failure of MOST over complex terrain, and especially under stable stratification. In addition, the scaling of streamwise $\Phi_{\varepsilon_u}$ and surface-normal $\Phi_{\varepsilon_w}$ dissipation rates differ, emphasizing that the prevalent anisotropy of complex atmospheric turbulence persists to the equilibrium range (59, 60), based on which the dissipation rates are determined.

Both the temperature variance and dissipation rates for unstable and stable stratification show a striking, albeit non-linear, dependence on turbulence anisotropy (Fig. 2d-f). As already observed for momentum variances, the seamless way that anisotropy incorporates into traditional MOST is through the constants in the traditional scaling relations, which also depend on anisotropy for these scaling variables (cf. Table 1). The non-linear nature of this dependence, though, prevents the derivation of a new normalized variable that would allow collapsing all the data onto a single curve, as was the case for the momentum variances (cf. Fig. 3). It also explains why Stiperski et al. (56) found no linear correlation between the degree of anisotropy and the departure from the respective canonical scaling curves for temperature variance. Still, the novel scaling relations (Fig. 3) offer a significant improvement to traditional scaling by reducing scatter on the corresponding family of curves that are dependent on turbulence anisotropy. Both temperature and dissipation rate scaling curves also satisfy the theoretical limits in free-convective regions (29). The novel scaling relations for TKE dissipation rates $\Phi_{\varepsilon_{u,w}}$ also show that the surface-normal dissipation rate is noticeably lower than the streamwise rate (cf. 40) for anisotropic turbulence in both stable and unstable stratification, and that this can be explained by bulk anisotropy.

**Improvement over traditional scaling.** To quantify the improvement of the new generalized scaling relations over the traditional MOST we define the skill score as

$$SS = 1 - \frac{MAD_{new}}{MAD_{MOST}}, \quad [3]$$

where $MAD$ is the median absolute deviation calculated as the departure of individual data points from the respective scaling curve $MAD = median(|x - x_{scaling}|)$ for each variable and stratification range. If the novel scaling outperforms traditional MOST (i.e. has lower scatter (16)), the SS will be positive.

Figure 4 show that the generalized scaling outperforms traditional MOST for all variables, for both stable and unstable stratification. The improvement is particularly obvious away from near-neutral conditions ($|\zeta| > 0.1$) where MOST has generally larger uncertainty. The comparatively marginal improvement in near-neutral conditions for some variables can be expected as MOST was traditionally developed for conditions that are closer to neutral.

Table 1. Functional form of the novel similarity scaling relations

| Scaling relation | Parametrization | Coefficients |
|---|---|---|
| **Unstable stratification** | | |
| $\Phi_u = a(y_B)(1-3\zeta)^{\frac{1}{3}}$ | $a(y_B) = \sum_{n=0}^{1} a_n (\log_{10} y_B)^n$ | $a_0 = 0.784, a_1 = -2.582$ |
| $\Phi_v = a(y_B)(1-3\zeta)^{\frac{1}{3}}$ | $a(y_B) = \sum_{n=0}^{1} a_n (\log_{10} y_B)^n$ | $a_0 = 0.725, a_1 = -2.702$ |
| $\Phi_w = a(y_B)(1-3\zeta)^{\frac{1}{3}}$ | $a(y_B) = \sum_{n=0}^{3} a_n (y_B)^n$ | $a_0 = 1.119, a_1 = -0.019, a_2 = -0.065, a_3 = 0.028$ |
| $\Phi_\theta = 1.07(0.05 + |\zeta|)^{-\frac{1}{3}} + (-1.14 + a(y_B)|\zeta|^{-\frac{9}{10}})(1 - \tanh 10|\zeta|^{\frac{2}{3}})$ | $a(y_B) = \sum_{n=0}^{1} a_n (y_B)^n$ | $a_0 = 0.017, a_1 = 0.217$ |
| $\Phi_{\varepsilon_u} = a(y_B)(1-\zeta)^{-1} - b(y_B)\zeta$ | $a(y_B) = \sum_{n=0}^{1} a_n (y_B)^n$<br>$b(y_B) = \sum_{n=0}^{2} b_n (y_B)^{-n}$ | $a_0 = 0.024, a_1 = 1.901$<br>$b_0 = 0.448, b_1 = 0.124, b_2 = 0.002$ |
| $\Phi_{\varepsilon_w} = a(y_B)(1-\zeta)^{-1} - b(y_B)\zeta$ | $a(y_B) = \sum_{n=0}^{1} a_n (y_B)^n$<br>$b(y_B) = \sum_{n=0}^{2} b_n (y_B)^{-n}$ | $a_0 = -0.059, a_1 = 1.844$<br>$b_0 = 0.263, b_1 = 0.086, b_2 = -0.004$ |
| **Stable stratification** | | |
| $\Phi_u = a(y_B)(1+3\zeta)^{c(y_B)}$ | $a(y_B) = \sum_{n=0}^{2} a_n (y_B)^n$<br>$c(y_B) = \sum_{n=0}^{4} c_n (y_B)^n$ | $a_0 = 2.332, a_1 = -2.047, a_2 = 2.672$<br>$c_0 = 0.255, c_1 = -1.76, c_2 = 5.6, c_3 = -6.8, c_4 = 2.65$ |
| $\Phi_v = a(y_B)(1+3\zeta)^{c(y_B)}$ | $a(y_B) = \sum_{n=0}^{2} a_n (y_B)^n$<br>$c(y_B) = \sum_{n=0}^{4} c_n (y_B)^n$ | $a_0 = 2.385, a_1 = -2.781, a_2 = 3.771$<br>$c_0 = 0.654, c_1 = -6.282, c_2 = 21.975, c_3 = -31.634, c_4 = 16.251$ |
| $\Phi_w = a(y_B)(1+3\zeta)^{c(y_B)}$ | $a(y_B) = \sum_{n=0}^{2} a_n (y_B)^n$<br>$c(y_B) = \sum_{n=0}^{4} c_n (y_B)^n$ | $a_0 = 0.953, a_1 = 0.188, a_2 = 2.253$<br>$c_0 = 0.208, c_1 = -1.935, c_2 = 6.183, c_3 = -7.485, c_4 = 3.077$ |
| $\log_{10}(\Phi_\theta) = a(y_B) + b(y_B)\log_{10}(\zeta) + c(y_B)\log_{10}(\zeta)^2 + d(y_B)\log_{10}(\zeta)^3$ | $a(y_B) = \sum_{n=0}^{1} a_n (y_B)^n$<br>$b(y_B) = \sum_{n=0}^{3} b_n (y_B)^n$<br>$c(y_B) = \sum_{n=0}^{3} c_n (y_B)^n$<br>$d(y_B) = \sum_{n=0}^{3} d_n (y_B)^n$ | $a_0 = 0.607, a_1 = -0.754$<br>$b_0 = -0.353, b_1 = 3.374, b_2 = -8.544, b_3 = 6.297$<br>$c_0 = 0.195, c_1 = -1.857, c_2 = 5.042, c_3 = -3.874$<br>$d_0 = 0.0763, d_1 = 1.004, d_2 = 2.836, d_3 = -2.53$ |
| $\log_{10}(\Phi_{\varepsilon_u}^2) = a(y_B) + b(y_B)\log_{10}(\zeta) + c(y_B)\log_{10}(\zeta)^2$ | $a(y_B) = \sum_{n=0}^{1} a_n (y_B)^n$<br>$b(y_B) = \sum_{n=0}^{3} b_n (\log_{10} y_B)^n$<br>$c(y_B) = \sum_{n=0}^{3} c_n (y_B)^n$ | $a_0 = 0.56, a_1 = 1.474$<br>$b_0 = 0.225, b_1 = -4.217, b_2 = -5.103, b_3 = -1.469$<br>$c_0 = -0.135, c_1 = 2.892, c_2 = -6.814, c_3 = 4.78$ |
| $\log_{10}(\Phi_{\varepsilon_w}^2) = a(y_B) + b(y_B)\log_{10}(\zeta) + c(y_B)\log_{10}(\zeta)^2$ | $a(y_B) = \sum_{n=0}^{1} a_n (y_B)^n$<br>$b(y_B) = \sum_{n=0}^{3} b_n (\log_{10} y_B)^n$<br>$c(y_B) = \sum_{n=0}^{3} c_n (y_B)^n$ | $a_0 = -0.288, a_1 = 2.417$<br>$b_0 = 0.687, b_1 = -0.678, b_2 = 0.383, b_3 = 0.447$<br>$c_0 = 0.027, c_1 = 1.81, c_2 = -5.103, c_3 = 4.167$ |

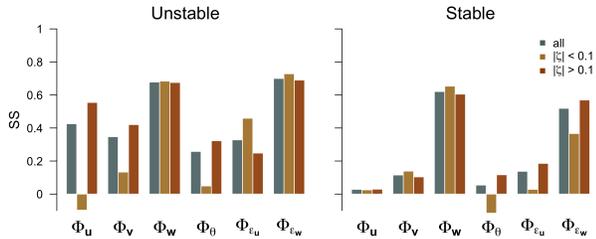

**Fig. 4.** Skill score measuring the improvement of the generalized scaling over traditional MOST for unstable (left) and stable (right) stratification, for scaled streamwise ($\Phi_u$), spanwise ($\Phi_v$) and surface-normal ($\Phi_w$) momentum, temperature ($\Phi_\theta$) and dissipation rates ($\Phi_\varepsilon$), for the entire stability range (blue), for near-neutral conditions (mustard) and strongly unstable/stable (burnt orange).

## Discussion

As MOST did at its inception, the novel scaling relations presented here provide a deeper insight into the fundamental nature of turbulence over both canonical and highly heterogeneous surfaces. The generalized scaling confirms the universal importance of turbulence anisotropy, as it is shown to impact not only the transport of momentum, but also the transport of scalars, and to allow successful scaling over vastly different terrain. Thus, the results suggest that understanding the impact of terrain and flow complexity on anisotropy is the key to explaining its effect on turbulence properties.

Let us first focus on unstable stratification. The momentum scaling shows clearest dependence on anisotropy in strongly convective conditions ($\zeta < -1$) where streamwise momentum $\Phi_u$ is lower for quasi-isotropic turbulence than anisotropic turbulence, and scales closer to the traditional surface-normal scaling curve (Fig. 2a-b). This signifies that quasi-isotropic turbulence carries overall less TKE (sum of all the variances) than anisotropic turbulence for the same stratification and the same momentum flux. The results also show that anisotropy ($y_B$) itself is not solely governed by strat-

Stiperski *et al.* June 20, 2022 | vol. XXX | no. XX | 5

ification, as all anisotropy types are observed for a range of $\zeta < -1$ (cf., 72). On the contrary, temperature variance shows no dependence on anisotropy in strongly convective conditions (Fig. 2d). This difference is hypothesized to occur due to the dissimilarity between momentum and scalar transport. From a fluids and heat transfer perspective it is often assumed that turbulence in the ABL transports all scalars (i.e. temperature, moisture, pollutants) similarly (73), and often this is also extended to include momentum (i.e. Reynolds Analogy). However, in the ABL this simple analogy only holds under neutral conditions, and significant differences in how momentum and heat are transported under convective conditions have been reported (74, 75). In high-Reynolds number turbulent flows, the structure of turbulence in the boundary layer changes substantially when evolving from a neutral stratification to a free convective regime, with turbulence characterized by hairpin vortices, trains of hairpin vortices and elongated coherent structures in the former, and convective cells and vertical plumes in the latter (e.g., 76). Under neutral conditions transport of momentum and heat are synchronized, with the same turbulent updrafts and downdrafts equally transporting momentum and scalars (75). Alternatively, with increasing thermal instability, this similarity breaks down, and motions transporting momentum and heat become unsynchronized (75). In fact, the transition of coherent structures from eddies with significant horizontal vorticity to larger thermal plumes is suggested here as the main cause of the dissimilarity between the transport of momentum and scalars, captured by the differences of how anisotropy affects scaling of momentum and heat. In near-neutral conditions, on the other hand, the rate at which $\Phi_\theta$ approaches infinity as $\zeta$ approaches zero is hypothesized to be caused by the existence of heterogeneity-induced finite temperature fluctuations that persist despite the vanishing heat flux as neutral conditions are approached (29, 66). This increase towards infinity, not accounted for in traditional MOST, depends on anisotropy in the generalized scaling. Given that more isotropic turbulence has a steeper approach, our results suggest that for the same heat flux, more isotropic turbulence is associated with larger temperature variance than more anisotropic turbulence. Thus, one may hypothesize that thermal heterogeneity is important in driving more isotropic turbulence in near-neutral conditions.

The novel scaling curves under stable stratification, on the other hand, highlight the importance of anisotropy in coupling the transport of momentum and temperature variances in very stable stratification. Anisotropy is generally considered to increase with increasing stratification (67), since strong stratification dampens the surface-normal momentum variance and turbulence becomes characterized by pancake-like structures (55, 77). This increase of stratification leads to the energy transfer from TKE (captured by $\Phi_w$) to Turbulence Potential Energy (TPE, captured by $\Phi_\theta$), as the Total Turbulence Energy is conserved (67). While it is generally assumed that the conversion between TKE and TPE is a function of stratification, our results (Figs. 2c-d) show that the coupling is in fact effectuated through anisotropy, and that for more anistropic turbulence TKE decreases while TPE increases (warmer colors) compared to the more isotropic conditions (colder colors). This inverse coupling is observed throughout a range of stratifications ($\zeta > 10^{-2}$), suggesting that neither $\zeta$ nor the Richardson number (not shown) are directly controlling anisotropy. At the same time, no dominant change in horizontal momentum variances is observed as a function of anisotropy, confirming that other processes (e.g. slope angle, wind turning with height, sub-mesoscale motions) play a more prominent role in their scaling (56). Finally, the clear dependence of dissipation rates on anisotropy confirm the findings of Stiperski et al. (59) showing a direct link between anisotropy at integral turbulent scales ($y_B$) and anisotropy at small scales embedded in the dissipation rates ($\varepsilon$).

## Conclusions

We present the first generalized scaling framework for complex atmospheric turbulence. The novel scaling relations based on turbulence anisotropy, build on the traditional Monin Obukhov Similarity Theory (MOST) used to parametrize surface-atmosphere turbulent exchange in all numerical weather prediction and climate models, but extend its limits of applicability to a comprehensive span of stabilities and a vast range of complex surface and flow conditions in which MOST is not valid.

The novel generalized scaling relations provide substantial improvement over traditional MOST. In particular, they address some of the most salient failures of MOST: the scaling of horizontal velocity variances in convective conditions, and scaling of surface-normal velocity variance, temperature variance and TKE dissipation rates in stable stratification. More importantly, the generalized scaling outperforms MOST not only for those datasets representing canonical conditions for which MOST was initially developed (i.e. horizontally homogeneous and flat terrain), but extends the scaling to cases that directly violate MOST requirements.

Additionally, the novel scaling relations allow a deeper insight into the fundamental nature of turbulence over both canonical and complex terrain. The critical role anisotropy plays in unifying the scaling over complex terrain is further proof that locally computed turbulence anisotropy encodes not only the imprint of large non-local convective eddies that are causing the failure of MOST in convective conditions, but also the complexity of surface characteristics and associated pressure perturbations found in mountainous terrain. The results also fit within existing theoretical frameworks used in describing atmospheric near-surface turbulence. Thus, anisotropy is shown to be the coupling mechanism between turbulent kinetic and potential energies in very stable stratification, as well as, accounts for the scalar dissimilarity in very unstable stratification.

Given the central role of MOST in parametrizing turbulence in all weather, climate, ocean and air pollution models, the new scaling presented here offers a major step forward in the improvement of turbulence representation over the majority of Earth's surface. This will help reduce uncertainty in climate projections of the most vulnerable locations, such as polar regions experiencing unprecedented rates of global warming, or mountainous regions undergoing elevation-dependent warming.

## Data and Methods

A detailed description of the datasets, turbulence post-processing procedures, anisotropy analysis, and scaling is given in Stiperski et al. (55, 56). Here we cover the salient aspects of methodology used.

**Data and Data processing.** We study data from 13 turbulence towers already investigated in previous studies (55, 56, 59, 72), chosen based on the requirement that a variety of terrain and flow conditions is represented, and a broad span of anisotropy is found in each tower's climatology. AHATS (62), CABAUW (63) and CASES-99 (61) are representative of canonical (horizontally homogeneous and flat terrain) conditions for which MOST is assumed to be valid. The Central and West tower from T-REX (64) and the NEAR and RIM towers from METCRAX II (65), can be classified as moderately complex terrain, as they are located on gentle slopes of up to 4 degrees. The exception is the METCRAX RIM tower located at the rim of the Meteor Crater, Arizona which, depending on wind direction, will be strongly influenced by very steep and complex topography of the rim. Finally, the five towers (CS-VF0, CS-SF1, CS-NF10, CS-NF27, CS-MT21) belonging to the i-Box long-term observational network (41) are located in highly complex alpine terrain at slopes of varying degrees (indicated by numbers in the tower name) over highly heterogeneous surfaces. Station CS-MT21, as well as an additional station (Im Hinteren Eis) are both ridge-top stations and therefore represent the most complex topographic and flow conditions. All datasets were processed with uniform processing to ensure their comparability. Data were rotated into the streamwise coordinate system in each averaging period using double rotation, where $u, v$, and $w$ correspond to streamwise, spanwise, and surface-normal velocity component. Data were de-trended prior to block averaging, while the averaging window length used to separate turbulence from non-turbulent motions depended on stratification and was 30-min for unstable, and 1-min for stable stratification. The dissipation rate $\varepsilon$ of turbulence kinetic energy was computed from the spectra of the streamwise ($\varepsilon_u$) and surface-normal ($\varepsilon_w$) velocity components using the inertial dissipation method (78), where the spectra were required to have a $-5/3$ slope in the inertial sub-range. Only data that were considered stationary within the averaging window according to the standard stationarity test (79) were considered. No other quality criteria were applied.

**Anisotropy Analysis.** Anisotropy of the Reynolds stress tensor in each averaging window was quantified through the anisotropy invariant analysis (57). As already employed in Stiperski et al. (55, 56), we follow the approach of Banerjee et al. (69) and define the degree of anisotropy (degree of departure of turbulence from isotropic conditions in which all turbulent variances are equal and all fluxes are zero) as $y_B = \sqrt{3}/2(3\lambda_3 + 1)$, where $\lambda_3$ is the smallest eigenvalues of the anisotropy stress tensor, defined by its components $b_{ij} = (\overline{u'_i u'_j}/\overline{u'_l u'_l}) - (1/3)\delta_{ij}$. Here $\overline{u'_i u'_j}$ are the components of the Reynolds stress tensor, and $\delta_{ij}$ is the Kronecker delta function. The extreme values of $y_B$ correspond to $y_B = 0$ for highly anisotropic (one- or two-component) turbulence, and $y_B = \sqrt{3}/2$ for purely isotropic turbulence.

**Scaling relations.** The new scaling relations are developed in the local scaling sense in the tradition of other scaling studies (30, 55, 56). This approach has the advantage of applicability both in unstable and stable stratification, and also allows scaling in complex terrain where fluxes vary with height (e.g. 56). In addition, we use an ensemble approach as suggested by Kader and Yaglom (29) and adopted in Stiperski et al. (56, 59). Data from all datasets were therefore combined into an ensemble, and then clustered according to the degree of anisotropy into ten clusters spanning the $y_B = [0.1 - 0.7]$ range, for unstable ($\zeta < 0$) and stable ($\zeta > 0$) stratification separately. Random sampling with replacement was used to ensure equal representation of each dataset in every anisotropy cluster. Bin medians of each scaled variable ($\Phi_u = \sigma_u/u_*, \Phi_v = \sigma_v/u_*, \Phi_w = \sigma_w/u_*, \Phi_\theta = \sigma_\theta/\theta_*, \Phi_{\varepsilon_u} = kz\varepsilon_u/u_*^3, \Phi_{\varepsilon_w} = kz\varepsilon_v/_*^3$), in each cluster were then computed for twenty logarithmically spaced $\zeta$ bins ($|\zeta| = [10^{-4} - 10^2]$). The functional form of the respective traditional MOST scaling relations for each scaled variable (cf., first column in Table 1) was then fit through the bins in each cluster, using non-linear robust fit. To reduce the uncertainty of the fit, we used bootstrapping and repeated the random sampling, binning, and fitting procedures a hundred times, so that the final fitting coefficients were calculated from the medians of the hundred best-fit estimates. Finally, the dependence of the fitting coefficients on anisotropy of each cluster ($a(y_B), b(y_B), c(y_B), d(y_B)$) was used to develop a polynomial parameterization (cf., second and third column in Table 1) incorporating the influence of anisotropy into the novel scaling curves.

**ACKNOWLEDGMENTS.** The work of Ivana Stiperski is part of a project that has received funding from the European Research Council (ERC) under the European Union's Horizon 2020 research and innovation programme (Grant agreement No. 101001691).Marc Calaf is thankful for the support of the National Science Foundation Grants PDM-1649067 and PDM-1712538, as well as the support of the Alexander von Humboldt Stiftung/Foundation, *Humboldt Research Fellowship for Experienced Researchers*, during the sabbatical year at the Karlsruhe Institute of Technology Campus Alpin in Garmisch-Partenkrichen.